\RequirePackage{fix-cm}
\documentclass[smallextended]{svjour3}       
\smartqed  
\usepackage{graphicx}
\usepackage{hyperref}
\journalname{Engineering Applications of Artificial Intelligence}

\begin{document}

\title{Scoring and Assessment in Medical VR Training Simulators with Dynamic Time Series Classification
}


\author{Neil Vaughan         \and
        Bogdan Gabrys 
}


\institute{N.Vaughan \at
              University of Exeter, Institute of Biomedical and Clinical Science, RILD Building, Barrack Road, Exeter, United Kingdom \\
              Tel.: +44-7783-527-327\\
              \email{n.vaughan@exeter.ac.uk}           
           \and
           B. Gabrys \at
              University of Technology Sydney, Advanced Analytics Institute, 15 Broadway, Ultimo NSW 2007, Australia
}

\date{Received: 21st August 2019 / Accepted: 8th June 2020}

\maketitle

\begin{abstract}
This research proposes and evaluates scoring and assessment methods for Virtual Reality (VR) training simulators. VR simulators capture detailed n-dimensional human motion data which is useful for performance analysis. Custom made medical haptic VR training simulators were developed and used to record data from 271 trainees of multiple clinical experience levels. DTW Multivariate Prototyping (DTW-MP) is proposed. VR data was classified as Novice, Intermediate or Expert. Accuracy of algorithms applied for time-series classification were: dynamic time warping 1-nearest neighbor (DTW-1NN) 60\%, nearest centroid SoftDTW classification 77.5\%, Deep Learning: ResNet 85\%, FCN 75\%, CNN 72.5\% and MCDCNN 28.5\%. Expert VR data recordings can be used for guidance of novices. Assessment feedback can help trainees to improve skills and consistency. Motion analysis can identify different techniques used by individuals. Mistakes can be detected dynamically in real-time, raising alarms to prevent injuries.
\end{abstract}

\section{Introduction}
\label{intro}
Virtual Reality (VR) training simulators are growing in popularity and are used by aircraft pilots \cite{yav}, surgeons and clinicians \cite{jud}, \cite{hua}, \cite{vaughan2014parametric}, military and defence applications \cite{kho}. Training simulators enable trainees to learn the skills required to perform skilled tasks by practicing on a virtual model in-vitro. The advantages of using simulators for surgery are well documented including the reduced risk of injury associated with practice on patients \cite{len}, ability to safely practice emergency procedures and ability to practice on various models of different patients \cite{vaughan2015}. Additionally, due to changing training structure and compliance with the European Working Time Directive, clinical experts are being required to reduce their time spent assessing novices \cite{wil}. Virtual training is useful in high risk activities such as epidural needle insertion, helping to avoid injuries \cite{gup}.
Automated training with VR simulators could assist, however methods for skill assessment in VR, particularly dynamically in real-time, are still undeveloped. The following sections outline recent methods for classifying skill and define our proposed method, which is applied and tested for accuracy.

\subsection{Background}
\label{sec:background}

Skill classification in surgical VR training has been attempted using various approaches \cite{vaughan2016}, including time series clustering and classification which received considerable research attention \cite{mon}, \cite{pap}. 
The JIGSAWS dataset \cite{gao} provides benchmark data from Da Vinci robot for testing skill classification and 100\% has been achieved using deep learning \cite{for} specifically for surgical skill assessment. Since 2018 this emerging field of Surgical Data Science rapidly accelerated with increases in Deep Learning for multivariate Time Series Classification. Recent advances producing state of art results in general time series classification include long short term memory (LSTM), a recurrent Fully Convolutional Network (FCN) for multivariate series \cite{kar} and TimeNet, a multilayered Recurrent Neural Network (RNN) \cite{mal}, inspired by successful image feature extraction.
Machine learning algorithms with inertial measurement units can improve the predictive power of surgeon motion analysis \cite{wat}. Support Vector Machine (SVM) classification can achieve 86\% sensitivity whereas the non-machine learning Lempel–Ziv (LZ) complexity metric gave 64\% sensitivity. This suggests that nonparametric supervised learning algorithm such as SVM applied to surgical skills classification can be useful for motion pattern recognition. \cite{oro} assessed skill for MIS, hand-eye bimanual coordination, spatial perception in a sample population of 4 experts, 22 residents, 16 novices, applying Linear discriminant analysis (LDA) (71\%), Nonlinear support vector machine (SVM) (78.2\%) and Adaptive neuro-fuzzy inference systems (ANFIS) (71.2\%). Fuzzy classification and radial bias function (RBF) was used \cite{hua} with MIST-VR dataset containing 4 experts, 4 intermediate, 4 novices over 200 Epochs gaining 33\%. Support vector machine (SVM) \cite{all} gained 91.6\%. Hidden Markov Models (HMMs) can be used to classify surgeon skills from surgical gestures with accuracy up to 100\% and discovers rules governing task ordering \cite{tao}.
Surgical skill can be classified using global measurements of the simulation, such as the distance travelled \cite{dat}, the total time taken \cite{jud}, force or pressure signatures \cite{yam}, the number and speed of hand motions \cite{dat}. These global measurements offer the quickest methods but lack information about task structure. Dynamic time warping (DTW) could further benefit classification.

\subsection{Dynamic Time Warping}
\label{sec:dtw}

Given \(\{a_n\}\) and \(\{b_n\}\) are two multivariate time series, various local distance functions, denoted \(\delta\), are compatible with DTW. Euclidean distance (Eq. 1) can be applied if both series are of equal length. Squared Euclidean distance uses the same formula (Eq. 1) without the square root, reducing computation. Manhattan city block is widely used and rapidly computed. Minkowski forms a generalisation of Euclidean and Manhattan distances. Others distance measures include Mahalanobis \cite{mah}, Bhattacharyya \cite{bha} and Canberra \cite{lan}.

\begin{equation}
\sum\limits_{i=1}^n\sqrt{\sum\limits_{v=1}^5(a_{vi}- b_{vi} )^2}
\end{equation}

where \(a_{vi}\) and \(b_{vi}\) refer to variable \(v\) from element \(i\) within \(\{a_n\}\) and \(\{b_n\}\), both containing \(n\) elements with \(5\) variables.

The dynamic programming algorithm DTW distance measure supports multivariate time series of unequal length where \(n \ne m\) \cite{ber}. A comparison between Euclidean and DTW is shown in Fig. 1.

\begin{figure}
  \includegraphics[width=1.00\textwidth]{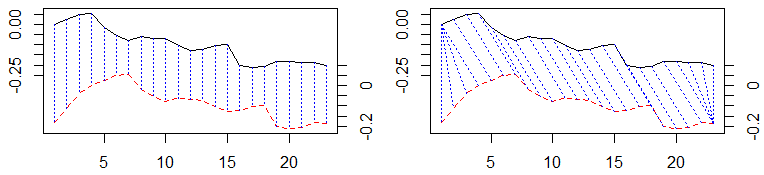}
\caption{Alignment by (left) Euclidean and (right) DTW distance measure. The upper time series is drawn vertically shifted to enhance visualization.}
\label{fig:1}       
\end{figure}

\begin{figure}
  \includegraphics[width=1.00\textwidth]{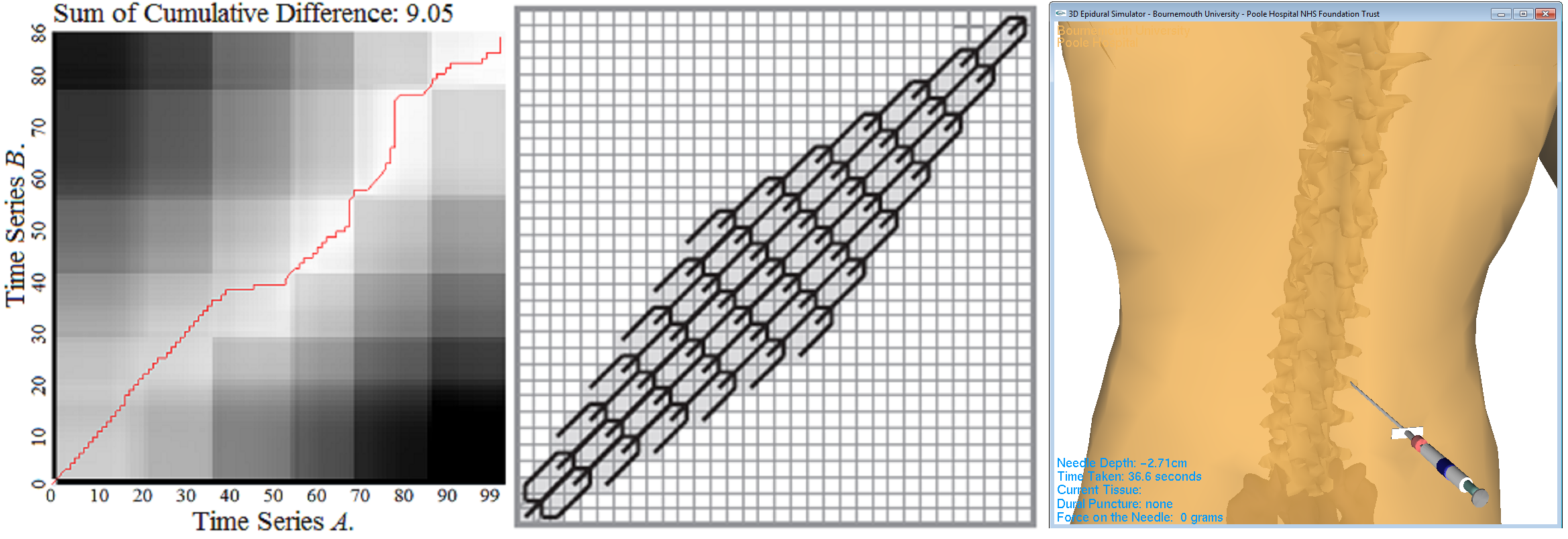}
\caption{(left) Local Cost Matrix (LCM) with warping path W shown as red line, (middle) Itakura Parallelogram, (right) developed epidural simulator.}
\label{fig:2}       
\end{figure}

Visualization of the DTW with warping path \(\{w_n\}\) is shown in Fig. 2, left. The Global constraints Itakura Parallelogram \cite{sak} applies (Fig. 2, middle) (Eq. 2).

\begin{equation}
w_i=w_{i-1}+min[(a_{n+1},b_{m+1}),(a_{n+2},b_{m+3}),(a_{n+3},b_{m+2})] 
\end{equation}

\subsection{Time Series Prototyping}
\label{sec:tsp}

Several time series clustering methods require combination of several time series into a single prototype time series, representing characteristics of all time series within a cluster \cite{gus}. Prototyping is an essential tool for many clustering algorithms like \(K\)-Means, or Ascendant Hierarchical Clustering to reposition cluster centroids and describe each cluster. Mean and median prototyping does not perform accurately and perturbs convergence of clustering algorithms, producing a non-representative prototype \cite{pet}, \cite{vaughan2016}. Partition Around Medoids (PAM) prototyping has a benefit that it avoids modifying any time series by calculating for each time series \(\{a_n\}\) the sum of distances to all other series, and selecting the prototype to be the one time series which has lowest total distance (Eq. 3).

\begin{equation}
PAM(\{a_n\})=DTW(\{a_n\},\{b_n\}) +  DTW(\{a_n\}, \{c_n\}) + DTW(\{a_n\},\{d_n\}) + ...
\end{equation}

DTW barycenter averaging (DBA) \cite{pet} is an iterative global method. The global nature of DBA enables the avoidance of iterative pairwise averaging, so results are unaffected by ordering. Shape based extraction \cite{pap} or fuzzy-based prototypes use fuzzy clustering such as fuzzy c-medoids (FCMdd) \cite{iza}. 
For this work, we propose a new method of time series prototyping: DTW Multivariate Prototyping (DTW-MP).

\subsection{Dynamic monitoring with Upper and Lower Envelopes}
\label{sec:dynamic}

Upper and lower envelopes are generated in this work to create a tunnel of acceptable motion using lower bounding so that an alert can be triggered if a VR object exits from the normal path of motion. Existing lower bounding methods have been proposed including: \(LB\_Yi\) \cite{yi}, \(LB\_Kim\) \cite{kim}, \(LB\_Keogh\) \cite{keo2005}, \(LB\_Improved\) \cite{lem}.
For this work there are three purposes of applying lower bounds: (1) To speed up the classification of new insertions by reducing the complexity of the similarity search required for new insertions, (2) To enable an alarm to be raised if the new time series does not stay within the tunnel of acceptable motion, between the upper and lower envelope around the cluster’s prototype insertion. (3) The upper and lower envelopes define an area which will contain all of the expert insertions, and our proposed \(DTW-MP\) prototype.

\subsection{Dynamic Assessment of Incomplete Series}
\label{sec:dynamicassessment}

Skill classification taking place during an insertion only has access to a partial time series, which is the first part of a surgical procedure, but not the end, because the remainder of the procedure has not yet been completed. 
This problem is related to time series sub-sequences \cite{rak2012}. The given time series \(\{a_n\}\) of length \(n\)  represents a sequence, \(\{a_n\}=a_1,a_2,...,a_n\). A subsequence \(\{q_m\}=q_1,q_2,...,q_m\), is a shorter region of length \(m\) from within \(\{a_n\}\)  which starts at any position \(i\) within \(\{a_n\}\) (whereby \(i\) is restricted such that \(i\leq (n-m)\), whereby \(n\geq m\geq 1\).
There are requirements which we aim to achieve when dealing with dynamic data: (1) Estimate during a procedure what proportion of the procedure has been completed. (2) Identify if the procedure is running fast or slow in comparison to the training insertions. (3) Compute the distance between a new partial trajectory and the cluster prototype.
Recent research on detection of unusual time series events refer to discord subsequences \cite{yan}, outliers, unusual, abnormal \cite{naf}, \cite{pok}, novel, deviant or anomalous time series subsequences \cite{keo2005hot}, \cite{lap}.
Our previous research has outlined methods for prediction of time series \cite{lemke}, \cite{rut}, which are particularly relevant in the context of the incomplete time series and could be applied to predict events which are likely to arise in the time series, which is a future work.
We apply comparison of incomplete time series in this work to enable dynamic monitoring of insertions in real-time.

\section{Methods}
\label{methods}

\subsection{Development of an Epidural Simulator}
\label{sec:develop}

We developed a virtual reality epidural training simulator using 3 Degrees Of Freedom (DOF) haptic input with force feedback and epidural pressure measurements (Fig. 2, right) \cite{vaughan2014parametric}. The 3D graphics model contains vertebrae with software and haptic based biomechanical models of soft tissues based on measurements from our clinical trial with obstetric patients of various Body Mass Index (BMI).
The data generated from the VR simulator consists of multivariate time series recording position, force and pressure over time. The 3D motion of the tool in x, y and z planes is recorded over time by the haptic device. The fourth dimension is force applied by the user, measured within the haptic device. The fifth dimension is pressure, measured within the syringe plunger using a custom wireless microcontroller system \cite{vaughan2014epidural}. 
Each epidural procedure tends to last around 20 seconds, during which the measurements were recorded at 500MHz, with 2 millisecond intervals. The resulting time series lengths are approximately 10,000 for each of the 5 measurements.

\subsection{Data Collection Trial}
\label{sec:data}

We recorded simulator training data using our VR epidural training simulator \cite{vaughan2014epidural}. Seven participants were in two groups: Group-C contained 3 medical trained NHS clinicians, each with varying experience of performing epidurals on real patients: ClinicianA had performed around 1000 real epidurals, ClinicianB had performed around 100 insertions and ClinicianC had performed around 20 epidurals. 
Group-N contained 4 non-clinicians who were not medically trained. Within Group-N, NonClinicianA had performed over 300 simulated epidurals, and the other 3 participants (NonClinicianB, NonClinicianC, NonClinicianD) had never performed real or simulated epidurals before. 
Skill labels (N=Novice, I=Intermediate, E=Expert) were assigned to each insertion based on the experience level, clinical background and number of clinical epidurals each clinician had previously completed.
In total 271 needle insertions were recorded. NonClinicianA recorded 101 epidurals. NonClinicianB, NonClinicianC and NonClinicianD recorded 95, 31 and 20 epidurals. ClinicianA recorded 10, ClinicianB recorded 9 and ClinicianC recorded 5.
To avoid data skew, epidural-40 subset of the 271 epidural dataset was created containing 40 insertions which exactly matches the class distribution and size of JIGSAWS dataset \(N, I, E, E, I, N, N, N\) with 8 participants performing 5 insertions each. This 40 epidural dataset was created at both lengths of 500 and 5000.
All of the data recordings were of simulated epidural insertion, using the same haptic device, the same build version of the VR software and on the same computer.
Recordings from 271 insertions were stored as multivariate time series \(\{a_n\}\) containing 5 variables: \(x, y, z, pressure, force\). Each series has different length, relative to the time taken.  

\subsection{Normalization of the Data}
\label{sec:normal}

Due to collection methods, standardization of the raw data recorded is necessary to set microcontroller sensor data onto the same scale as the haptic device data. 
A standard normalization method is applied which scales each element \(a_n\) within the time series \(\{a_n\}\) by subtracting the population mean from \(a_n\) and diving by the standard deviation \(\sigma\).
After normalization if one time series contains very subtle movement on \(z\) axis and another has large movement on \(z\) axis, these will be normalized into the same scale.

\subsection{Skill classification Method 1: DTW 1-NN}
\label{sec:skill}

The DTW-\(k\)-NN classifier was used to classify insertion skill, making use of all recorded insertions. Each of the training examples is checked to identify which most closely resembles the new insertion, by calculating the DTW distance between the new insertion and all previous recorded examples. The new insertion is labelled the same class as the closest training example (N, I or E).

\subsection{Skill classification Method 2: Nearest Centroid}
\label{sec:skill2}

Nearest centroid classifier applies the \(k\)-NN classifier by measuring DTW distance between the new insertions and the cluster prototypes, reducing the number of DTW computations required compared to DTW-\(k\)-NN which requires all insertions. 
A range of 7 state-of-art prototype methods were applied: Mean, SoftDTW, DBA, PAM, Shape Extraction, and we propose 2 new prototype methods: DTW-MP\(_D\) and DTW-MP\(_I\).
The prototype of each skill level (N,I,E) is calculated to represent a prototypical insertion for each cluster. The prototypical insertions are built from all insertions in each cluster.
We propose the DTW Multivariate Prototyping (DTW-MP) algorithm to produce prototypical time series using DTW \cite{vaughan2016}. Our proposed prototyping method (DTW-MP) has advantages over Mean and Median: (1) DTW-MP retains features which occur in two time series at different times which would not be aligned by Mean. (2) DTW-MP can handle two series of different length unlike Mean. (3) The DTW-MP prototype is guaranteed to stay within the summative-envelope.
Our proposed prototyping method (DTW-MP) starts by calculating the warping path \(\{w_n\}\) between two time series (Fig. 2, left). The new DTW-MP prototype \(\{p_n\}\) is created with the same length as the warping path \(\{w_n\}\). Each element \(p_n\) in \(\{p_n\}\) is set to the mean of the two elements from \(\{a_n\}\) and \(\{b_m\}\) which were aligned in the equivalent element \(w_n\) of the warping path \(\{w_n\}\). Therefore, the length \(k\) of the new DTW-MP prototype \(\{p_n\}\), will be the same length as the warping path \(\{w_n\}\), which is \(max(m,n) \leq k \leq m+n\).

\subsection{Skill classification Method 3: Deep learning}
\label{sec:skill3}

Four deep learning techniques were used for time series classification: (1) a relatively deep Residual Network (ResNet) with 9 convolutional layers and a Global Average Pooling (GAP) layer \cite{he}. (2) Fully Convolutional Network (FCN) \cite{lon} with a final layer of Global Average Pooling (GAP). (3) Convolutional Neural Network (CNN) \cite{zha} with final discriminative layer taking the result of the convolutions to give probability distribution over class variables. (4) Multi-Channel Deep Convolutional Neural Network (MCDCNN) \cite{zhe} with architecture of a traditional deep CNN, plus the convolutions are applied in parallel on each dimension of the input MTS. These four DL architectures were chosen to provide a range of frameworks due to their previous successful applications to time series classification tasks. The DL implementation architectures were matching with the open source time series classification framework \cite{faw}. Each method was applied to the 40 epidural subset for skill classification.

\begin{figure}
  \includegraphics[width=1.00\textwidth]{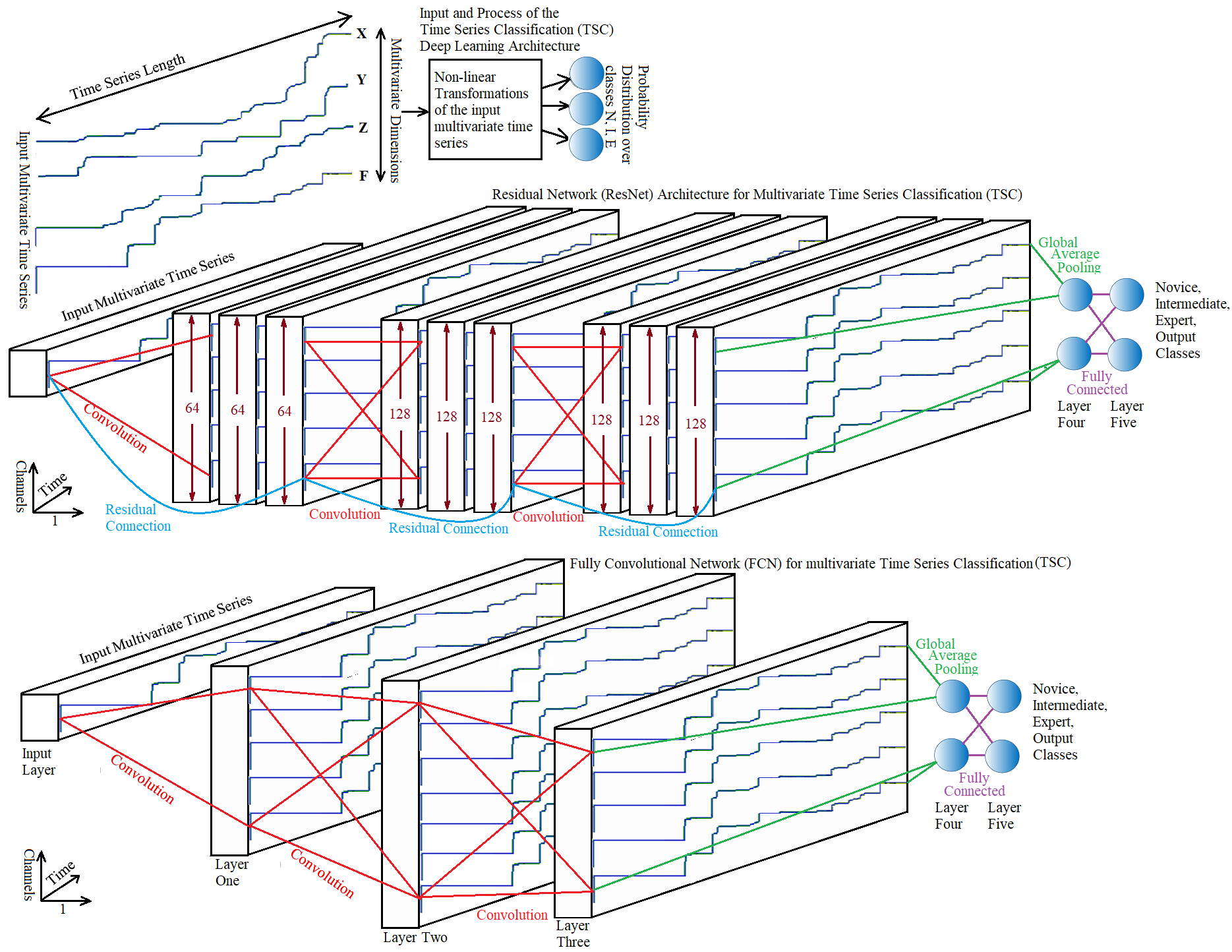}
\caption{Deep Learning Architecture for Multivariate Time Series Classification. (Top) input formatting process, (middle) ResNet Architecture, (lower) FCN Architecture.}
\label{fig:3}       
\end{figure}

Details of each deep learning architecture are in Fig. 3. Our method harnesses transfer learning, within FCN architecture, as one advantage of the utilized FCN method is the invariance which enables the use of a transfer learning approaches to train models on one dataset and further tune it on other target datasets. All the convolutions in the framework have a stride of 1 which preserves the length of the time series after convolution.
The methods we used to incorporate the time series data into the deep learning architectures is shown in Fig. 3 (top). The architectures for ResNet (Fig. 3, middle) and FCN (Fig. 3, lower) are shown. Architectures were based on adopted highest performing frameworks \cite{faw}. This illustrates that deep learning with multivariate time series is a challenging problem.

\section{Methods for Dynamic Monitoring}
\label{methodsdynamic}

\subsection{Phase estimation to detect proportion of time series}
\label{sec:phase}

It is first required to dynamically estimate how much of the procedure has been completed at any given time. Phase estimation is achieved by producing 10 subsequences of the prototype increasing in size from 10\% to 100\% at 10\% intervals. The DTW distance is computed between the incomplete new insertion and each of the 10 prototype subsequences. The best match identifies the proportion of the procedure which is completed, irrespective of whether one time series occurred faster than the other.

\subsection{Cluster Prototyping}
\label{sec:cluster}

Hierarchical clustering is applied to group similar good or bad insertions techniques together. In order to perform clustering, a distance matrix comparing each time series to all others is generated.
The DTW Euclidean normalised distance is pre-calculated between all pairs of insertions to produce a distance matrix of size \(271^2\), which requires \(\frac{271^2-271}{2}\) DTW computations. Within each time series \(\{a_n\}\), the number of elements \(a_n\) is approximately 1000, so each DTW comparison requires approximately \(1000^2\) element alignments which equates to \(36 \times 10^9\) Euclidean distance calculations. The distance matrix is used as input to hierarchical clustering.

\subsection{Upper and Lower Bounds tunnel of acceptable motion}
\label{sec:upper}

To enable dynamic real-time monitoring a tunnel or envelope of acceptable motion is created.
We propose summative-envelope algorithm which creates an envelope pathway guided in shape by the closest insertions to the expert cluster prototype. It generates a new pair of time series which contain a summative upper \(\{su_n\}\) (Eq. 4) and summative lower \(\{sl_n\}\) (Eq. 5) envelopes from the expert insertion envelopes. Each element of the summative upper envelope \(\{su_n\}\) is set to the maximum of any expert upper envelope at that time (Eq. 5), denoted as \(\{e1_n\}\), \(\{e2_n\}\), \(\{e3_n\}\), ... The summative-envelope is generated for all 5 variables in the multivariate time series, so the summative-envelope could be visualised as a 5D tunnel of acceptable motion.

\begin{equation}
su_{vi}= max(e1_{vi}, e2_{vi}, e3_{vi}, ...)
\end{equation}
\begin{equation}
sl_{vi}= min(e1_{vi}, e2_{vi}, e3_{vi}, ...)
\end{equation}

where \(su_{vi}\) refers to variable \(v\) from element \(i\) within time series \(\{su_{n}\}\).

This has benefit that new insertions will fit within the summative-envelope if they are within the envelope of any previously expert insertion, but not if it contains an event unlike any previously seen event in an insertion. 
The strength of our proposed summative-envelope technique is that the summative-envelope combines data from all insertions, whereas the \(LB\_Keogh\) lower bounding upper and lower envelopes only contain information from one time series. The summative-envelope only needs to be computed once from the training data. It can subsequently be used to raise an alarm each time a new insertion goes outside of the summative-envelope. During a simulation, summative-envelopes could enable the allowable motion to be visualised by the user in real-time, providing more clarity of procedural requirements. When reviewing completed simulations, the summative-envelope can enable a visualisation showing where and when it went wrong. If a new insertion is one of the closest to the prototype, the envelope is then re-generated when the insertion completes, adapting to the new data. This was previously a problem with black-box skill classification.

\subsection{Scoring a procedure based on tunnel of acceptable motion}
\label{sec:scoring}

The next step is to score each new insertion by checking that it remains inside a tunnel or envelope of acceptable motion which is created using prototypical insertion data. 
In order to score a new insertion dynamically during the insertion, the summative envelope from the identified cluster is used. The new insertion is monitored dynamically to identify whether it stays inside the envelope. If an insertion goes out of the envelope, an alarm is raised. The score is updated dynamically using Eq. 6, during the insertion taking into account what proportion of the insertion was outside of the envelope and the distance outside the envelope.

\begin{equation}
D(\{a_n\})=\sum\limits_{i=1}^n\sqrt{\sum\limits_{v=1}^5 max(a_{vi}- su_{vi}, sl_{vi} - a_{vi}, 0 )^2}
\end{equation}

Where \(D(\{a_n\})\) is the total sum of distances the new insertion \(\{a_n\}\) was outside the summative upper \(\{su_n\}\) and summative lower \(\{sl_n\}\) envelopes, which have 5 dimensions: x, y, z, pressure and force.

\section{Results for Skill Classification}
\label{results}

This section describes the application of the proposed methods to our VR simulator for epidural needle insertion.

\subsection{Skill Classification 1: DTW-1-NN}
\label{sec:skillclass}

DTW-1-NN was applied to classify skill of the Epidural-40 subset into classes N, I, E. Accuracy of DTW-1-NN was 60\%. 1-NN outperformed \(k\)-NN with \(k\) = 2 to 9. Results were verified by LOOCV and 5 fold CV. Accuracy wasn’t affected by length reduction from 5000 to 500, which doesn’t largely affect DTW distances \cite{rat}, \cite{vaughan2016}. The DTW-\(k\)-NN classification was also applied to the full dataset of 271 epidural insertions giving 90.03\% accuracy.

\subsection{Skill Classification 2: Nearest Centroid}
\label{sec:skillclass2}

Results from the nearest centroid classifier applied to the 40-epidural subset for skill classification showed that accuracy depends largely on the algorithm used for generating the centroid. Of the 7 prototype algorithms applied, SoftDTW gave highest accuracy (77.5\%) for skill classification, results for each are in Table I. The cluster prototypes are shown in Fig. 4. SoftDTW (red), Partition Around Medoids (PAM) (pink), DTW Barycenter Averaging (DBA)(Yellow), Shape Extraction (SE) (cyan), DTW-MP\(_I\) (green), DTW-MP\(_D\) (navy), Mean (black), and individual time series (grey) in each class (N, I, E).

\begin{table}
\caption{Nearest Centroid - Skill Classification Accuracy}
\label{tab:1}       
\begin{tabular}{lllllll}
\hline\noalign{\smallskip}
SoftDTW & Mean & DBA & SE & PAM & DTW-MPD & DTW-MPI \\
\noalign{\smallskip}\hline\noalign{\smallskip}
77.5\% & 70\% & 47.5\% & 60\% & 60\% & 50\% & 52.5\% \\
\noalign{\smallskip}\hline
\end{tabular}
\end{table}

\begin{figure}
  \includegraphics[width=1.00\textwidth]{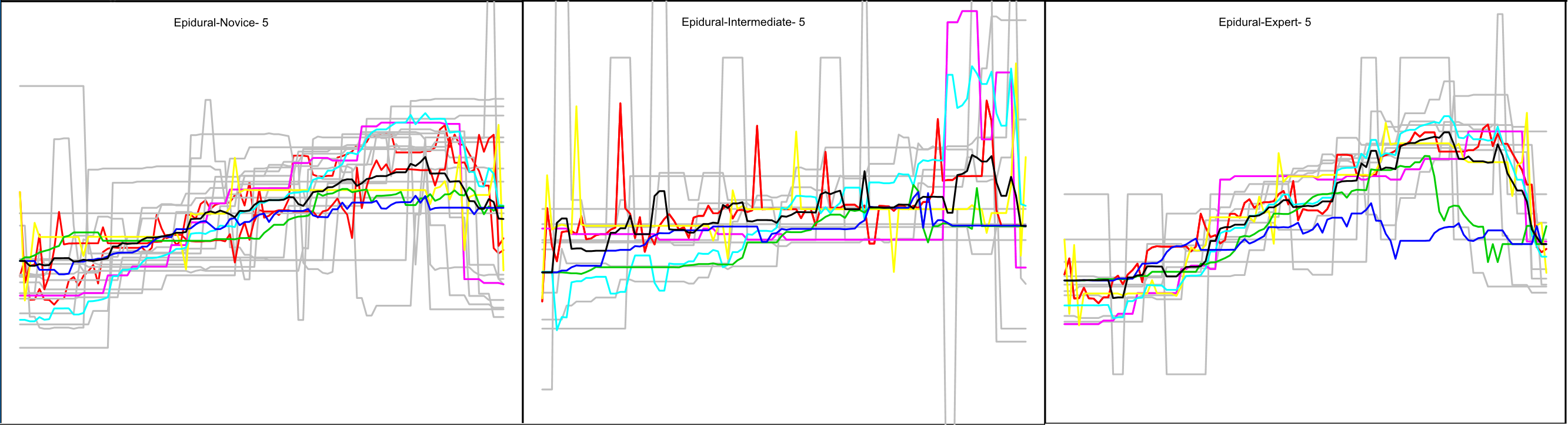}
\caption{All centroids - dimension 5 (Pressure) of Epidural-40 left-right: N, I, E}
\label{fig:4}       
\end{figure}

\subsection{Skill Classification 3: Deep Learning}
\label{sec:skillclass3}

Four deep learning techniques for time series classification were applied for skill classification. Time series length of 5000 and 500 were tested and produced similar results. Results are shown in Table II validated with both LOOCV (and 5-fold CV in brackets).

\begin{table}
\caption{Deep Learning - Skill Classification Accuracy}
\label{tab:2}       
\begin{tabular}{llll}
\hline\noalign{\smallskip}
ResNet & FCN & CNN & MCDCNN  \\
\noalign{\smallskip}\hline\noalign{\smallskip}
85\% (60.2\%) & 75\% (82.5\%) & 72.5\% (72.5\%) & 28.5\% (23.6\%) \\
\noalign{\smallskip}\hline
\end{tabular}
\end{table}

\section{Results From Dynamic Assessment}
\label{resultsdynamic}

\subsection{Clustering epidural insertions}
\label{sec:clustering}

Clustering was applied to the 271 insertions which produced seven clusters, as shown in the dendrogram in Fig. 5. This identifies the optimum separation between clusters and the optimum similarity within clusters according to the Cluster Validity Indices (CVIs). Most of the clusters contain a mixture of skill levels and individuals (Fig. 6). One individual can use several techniques or one technique can be used by several individuals. Clusters may represent numerous valid techniques of performing a good insertion, or common mistakes repeated by different people. For these reasons, when clustering was applied to produce 3 clusters representing each skill level (N, I, E), insertions from each skill level were not all grouped together.

\begin{figure}
  \includegraphics[width=1.00\textwidth]{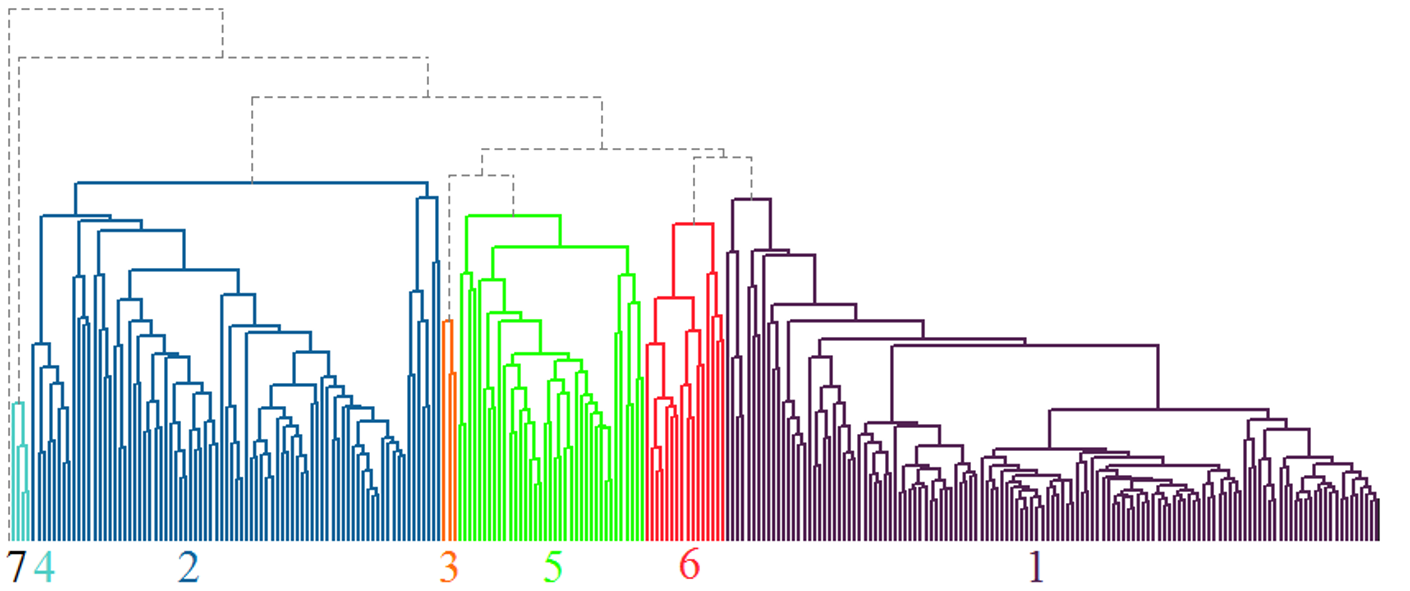}
\caption{Dendrogram of hierarchical clustering for 271 time series in 7 clusters}
\label{fig:5}       
\end{figure}

\begin{figure}
  \includegraphics[width=1.00\textwidth]{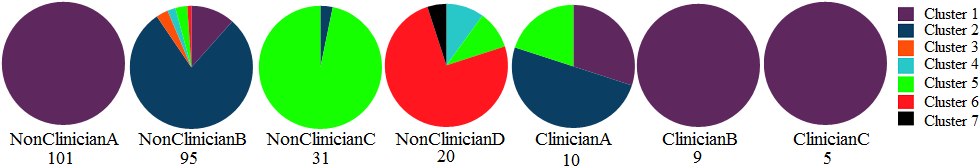}
\caption{The proportion of each cluster made up by each individual}
\label{fig:6}       
\end{figure}

Assessment of cluster validity was performed using seven CVIs including Sil, Dunn, COP, DB, DBStar, SF and CH \cite{arb}. The CVIs showed that highest validity was achieved by hierarchical clustering.  Generating seven clusters produced various numbers of insertions in each cluster: 129, 81, 3, 4, 37, 16, 1 with hierarchical clustering or 45, 83, 11, 24, 76, 26, 6 with partitional clustering.

\subsection{Prototypes of the 7 clusters}
\label{sec:prototypes}

A prototype was generated for each of the 7 clusters identified. Fig. 7 shows a comparison between the 7 prototypes on the \(x\) axis. This reveals why certain insertions were clustered separately. Our proposed prototyping method (DTW-MP) was used.

\begin{figure}
  \includegraphics[width=1.00\textwidth]{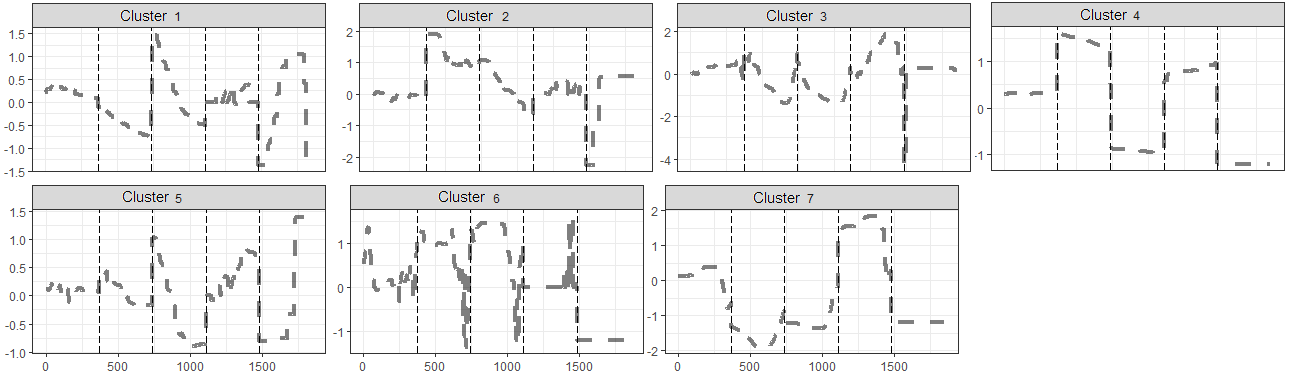}
\caption{The prototype of the \(x\) axis for each of the 7 clusters found using hierarchical clustering using DTW distance}
\label{fig:7}       
\end{figure}

\subsection{Scoring by distance from cluster prototype}
\label{sec:scoringby}

All insertions from cluster 1 are shown in Fig. 8. Highlighted black is the cluster 1 prototype insertion. Highlighted in Purple is the number one discord found in all trajectories which was furthest away from the prototype and highlighted in green is the sequence closest to the prototype, based on DTW with Euclidean distance.
The recorded trajectories and centroid was plotted into the VR simulator as shown in Fig 7, showing all trajectories in cluster 1. This is useful for the VR trainee who can visualise their performance in comparison to previous insertions.

\begin{figure}
  \includegraphics[width=1.00\textwidth]{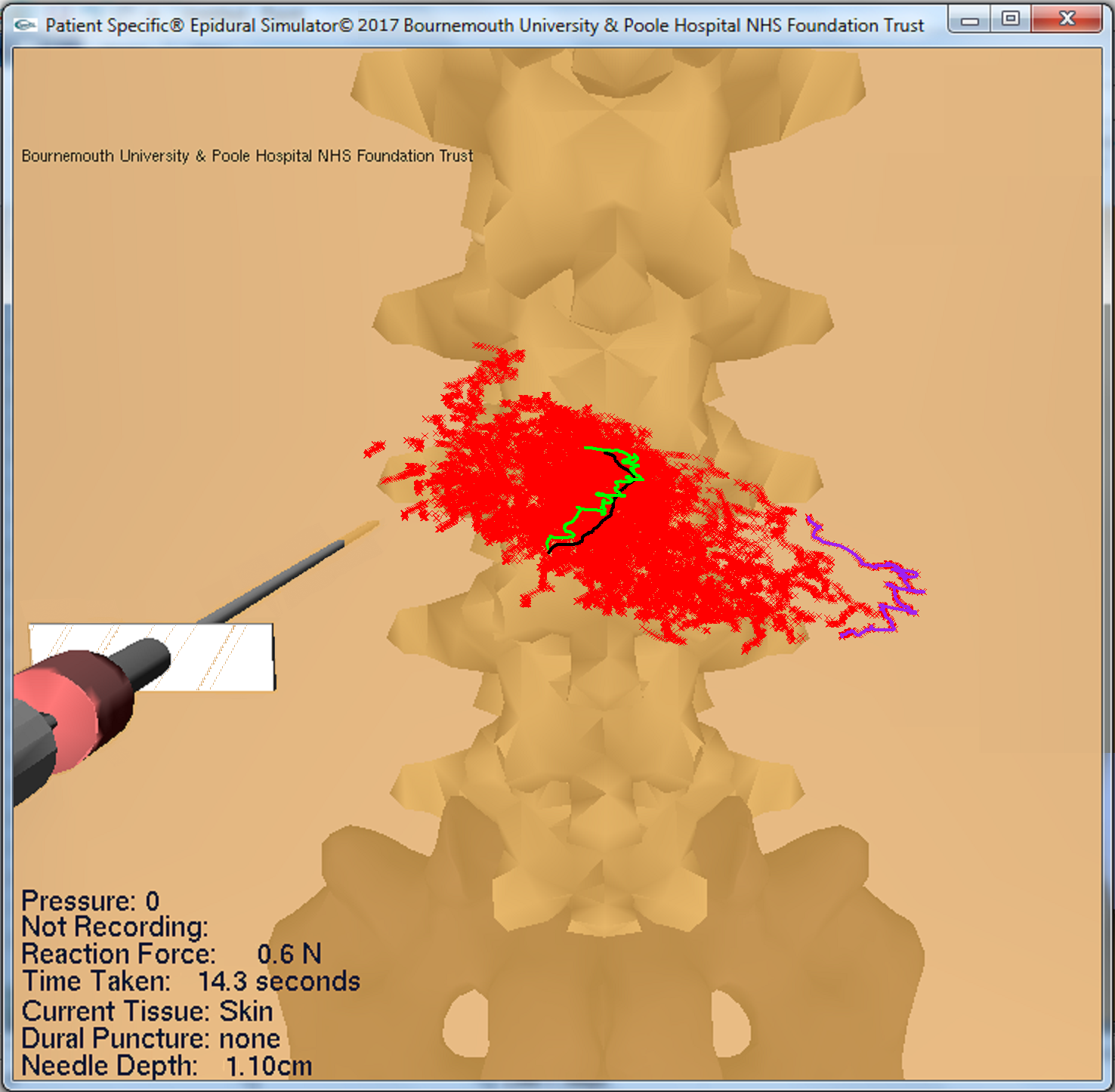}
\caption{Plot of all movement points from insertions in cluster 1. Black: The ideal prototypical trajectory. Green: The best insertion matching closely to the prototype. Purple: The worst insertion furthest from the prototype.}
\label{fig:8}       
\end{figure}

Each insertion can be dynamically scored according to the DTW distance from the prototype, generating the cumulative error cost matrix (Fig. 9). Cumulative error is much lower in the best trajectory (left) and higher error in the worst trajectory (right). This graph could be useful for clinicians to visualise which stage of the procedure contained most error. In the cost matrices (Fig. 9), the worst insertion has total cost of 656 whereas the best insertion has lower cost of 112 (shown on \(z\) axis in leftmost corner). This indicates that the worst insertion was approximately 5.8 times further from the ideal trajectory.

\begin{figure}
  \includegraphics[width=1.00\textwidth]{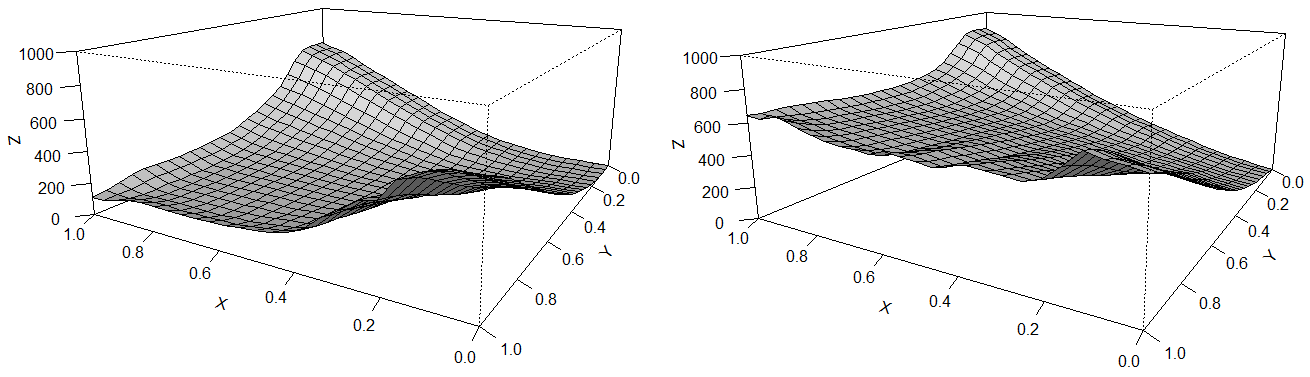}
\caption{Cumulative cost matrix of (left) best trajectory (right) worst trajectory.}
\label{fig:9}       
\end{figure}

\subsection{Generating envelopes for dynamic score monitoring}
\label{sec:generating}

The upper and lower bounds were generated from the prototypes of each cluster. \(LB\_Keogh\) is not compatible with multivariate data, so the upper and lower envelopes \(\{u_n\}\) and \(\{l_n\}\) were generated for each axis individually (Fig. 10).

\begin{figure}
  \includegraphics[width=1.00\textwidth]{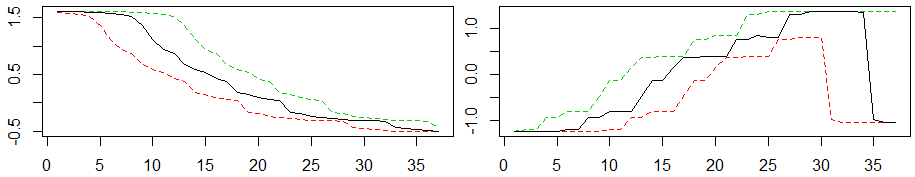}
\caption{The upper \(\{u_n\}\) (green) and lower \(\{l_n\}\) (red) envelopes for (left) z axis and (right) pressure from the prototype of cluster 1 using \(LB\_Keogh\) lower bounds.}
\label{fig:10}       
\end{figure}

The summative-envelopes were computed around the prototypes of each cluster, including the 4 best insertions in each cluster, according to their DTW distance from the cluster prototype. In Fig. 11, \(\{su_n\}\) and \(\{sl_n\}\) were created by taking the four insertions (solid black lines) which were closest to the prototype of cluster 1. Then \(LB\_Keogh\) lower bounding algorithm was applied to generate the 4 upper envelopes (dashed red lines) and 4 lower envelopes (dashed green lines). The envelopes were combined by creating the summative-envelope for cluster 1 shown in Fig. 11 as bold lines. 
Two unseen insertions (solid purple lines) are shown, which both fit within the summative envelope, so they would be classified as good insertions.
Where subsections of the insertion require higher accuracy than others, the summative-envelope intuitively produces a thinner tunnel in those areas.

\begin{figure}
  \includegraphics[width=1.00\textwidth]{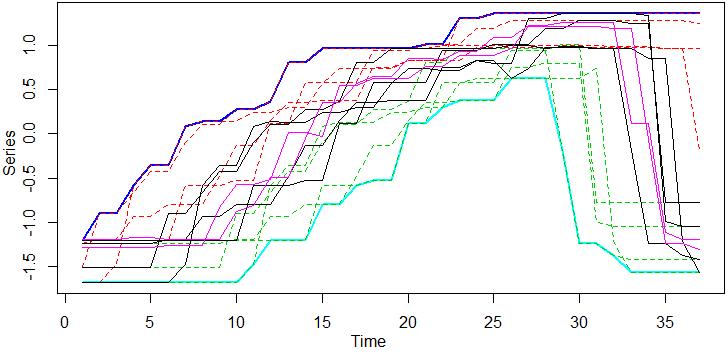}
\caption{Proposed Summative-envelope algorithm (bold lines) for syringe pressure around 4 insertions closest to the prototype of cluster 1.}
\label{fig:11}       
\end{figure}

\subsection{Phase estimation with incomplete time series}
\label{sec:phaseest}

The phase estimation algorithm predicted correctly 90\% of the time which proportion of the insertion had taken place. Even with a bad insertion, this method can detect that it’s a bad insertion even with only the first 10\% of the data, as the normalised distance is already much greater than that of the best insertions. During insertion, Fig. 12 shows that the completed part of an incomplete insertion has low normalised distance when compared to a similar percentage of the prototype, but a high distance when compared to a different percentage of the prototype. The percentages within Fig. 12 indicate the current percentage that has been completed of an incomplete insertion.

\begin{figure}
  \includegraphics[width=1.00\textwidth]{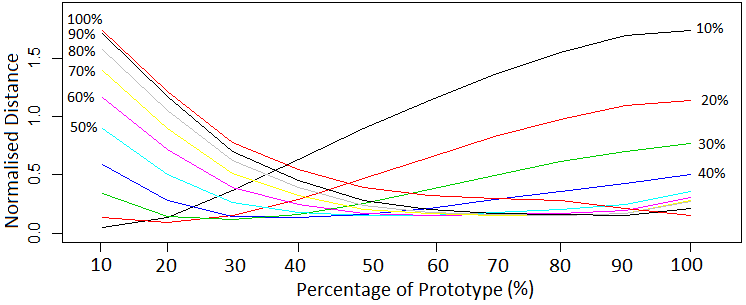}
\caption{Dynamic comparison between a partial time series during insertion and partial prototypes of various window length.}
\label{fig:12}       
\end{figure}

\subsection{Dynamic scoring}
\label{sec:dynscore}

The dynamic scoring begins by computing the distance between a new incomplete insertion and equal proportion of the prototype predicted by phase estimation. This results in a dynamic system able to score a partial insertion in real-time during insertion. Fig. 13 shows the dynamic scoring process applied to four new unseen partial insertions (green). All four new insertions stayed within the summative-envelope upper (dark blue) and summative-envelope lower (light blue) but some were closer to the cluster prototype (black) than others. The envelope has differing width in some parts, which represents the variation in accuracy required in some stages of the procedure.

\begin{figure}
  \includegraphics[width=1.00\textwidth]{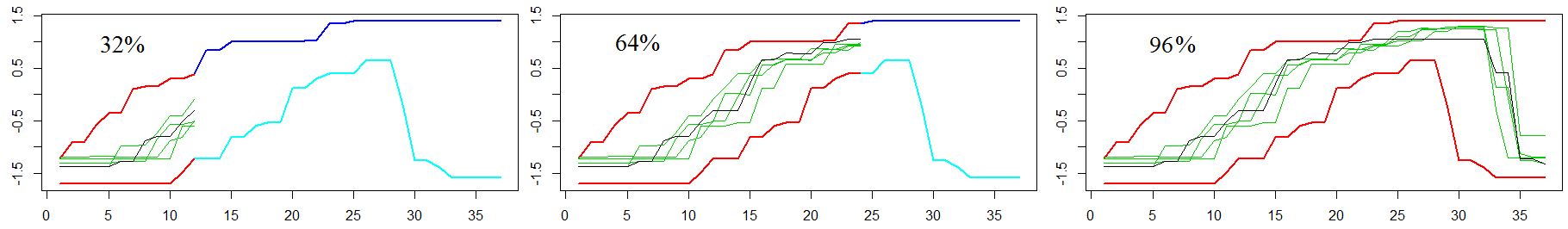}
\caption{Dynamic scoring of syringe pressure: partial new unseen time series shown in green, with the corresponding proportion of the Cluster 1 prototype in black. The Cluster 1 summative-envelope upper in dark blue, lower in light blue, turning red at the corresponding proportion.}
\label{fig:13}       
\end{figure}

The system can offer dynamic adaptive learning by adding new insertions into the training set. When a new insertion is combined with an existing cluster, it is possible either to re-compute the prototype including the new insertions or combine the new insertion with the existing prototype with weighting. Unusual new insertions can be identified if they are allocated into existing clusters with known bad insertions. These can then be added into the training set as bad insertions.

\subsection{Identifying the individual}
\label{sec:ident}

The DTW-\(k\)-NN classifier was applied to identify who performed each insertion. Of the 247 non-clinician insertions 95.4\% were classified as non-clinicians. Some individuals were easier to classify than others, NonClinicianA was correctly classified 100\% of the time due to high consistency (Fig. 14), whereas ClinicianA only 20\%. Overall, the individual was classified correctly in 81.2\% of insertions.

\begin{figure}
  \includegraphics[width=1.00\textwidth]{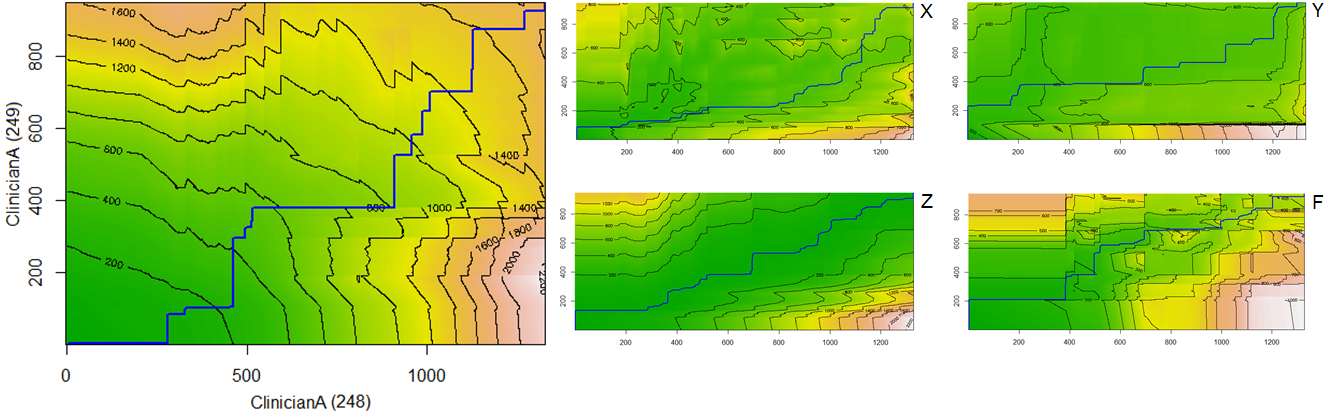}
\caption{Left: DTW comparison between two of ClinicianA’s insertions. Right: DTW comparison between the same two insertions using univariate data for \(x, y, z\) and \(force\) dimensions individually.}
\label{fig:14}       
\end{figure}

\section{Discussions}
\label{discussions}

This research proposed three methods for skill classification by analyzing multivariate time series recorded during a VR epidural training simulator procedure. Our data collection of 271 virtual reality epidural procedures included three trained clinicians from the NHS. (1) DTW-1-NN achieved 60\% classification accuracy. (2) Nearest centroid classifier SoftDTW achieved 77.5\%. (3) Within deep learning methods, ResNet achieved 85\%, FCN 75\%, CNN 72.5\% and MCDCNN was inaccurate with 28.5\%. Therefore the nearest centroid approach is competitive and only outperformed by the ResNet deep learning architecture. High performance of ResNet matches with previous TSC results applying DL methods (Fawaz et al., 2019).
Insights into why SoftDTW was the optimum prototyping method for nearest centroid include that SoftDTW is differentiable and both its value and gradient can be computed with quadratic time/space complexity making it more suited to cluster time series under the DTW geometry. In this case ResNet outperformed SoftDTW leveraging favorable features of our 5-dimensional multivariate time series dataset. Our dataset originated from epidural needle procedures which are relatively short compared to DaVinci surgeries which commonly produce longer, higher dimensional time series and future work could investigate whether SoftDTW or ResNet would perform similarly on those additional datasets. 
We proposed a new time series prototyping algorithm, DTW-MP which is applied to create a prototype insertion for each cluster. New insertions are classed according to their DTW distance from the cluster prototype. 
The research developed dynamic methods to assess the score of a virtual reality task while the task is being completed. (1) Clustering is performed to divide the time series training set into groups according to the different techniques. (2) We propose the Summative-envelopes algorithm, which takes the best time series from each cluster including the prototype to create a combined envelope using lower bounds. This can raise alarms in real-time if a time series exits the envelope tunnel. 
Our experiment showed that DTW-1-NN can recognise which trainee performed a virtual reality task in 81\% of the cases.
The developed methods enable trainees to view their score, clustering and summative-envelope in real-time during insertion. The summative-envelope reveals which parts of the procedure were abnormal. After reviewing the performance, the trainee’s technique can improve by repetitive practice until their motion becomes closer to the cluster prototype representing an expert, improving performance and skill of trainees. Over time, trajectory clustering algorithms can enable the measurement of consistency within a single trainee’s performances, and identify the trainee’s improvement of consistency over time.
Future work can use these results for several purposes including: (i) Adaptation and automation of VR training based on the recorded data to customise VR training for individual requirements. (ii) Detecting the type of motion or hand gestures using classification. (iii) Recognising actions which increase the risk of injury to raise an alert. 
In future, the developed methods could be applied to in-vivo data, tracking devices or cameras monitoring surgeons with hospital patients as well as being applied to trajectories from VR training simulators.

\section{Ethics}
\label{ethics}

The Bournemouth University Ethics service has reviewed the study plan prior to initiation. The dataset does not contain identifying information such as names or personal information.

\begin{acknowledgements}
This project was supported by the Royal Academy of Engineering (RAEng) under the Research Fellowship scheme awarded to Professor Neil Vaughan, also support from University of Exeter, University of Technology Sydney and Bournemouth University during the time of the research.
\end{acknowledgements}

%
\section*{Conflict of interest}
The authors declare that they have no conflict of interest.

\bibliography{mybib}
\bibliographystyle{spmpsci}      

This manuscript version is the author's accepted manuscript without typesetting $\copyright$ 2020. This manuscript version is made available under CC-BY-NC-ND 4.0 license \url{http://creativecommons.org/licenses/by-nc-nd/4.0/}

Please cite this article as: N. Vaughan and B. Gabrys, Scoring and assessment in medical VR training simulators with dynamic time series classification. Engineering Applications of Artificial Intelligence (2020) 103760, \url{https://doi.org/10.1016/j.engappai.2020.103760}

\end{document}